# Uncertainty Relations in Self-Similar Convergent Trajectories

Héctor Ceceña-Álvarez., Antonio Peimbert-Torres., Raúl W. Gómez-González.

**Abstract**: The Koch curve is a self-similar object whose length grows unboundedly when the measuring unit by which is calculated diminishes. If this curve is considered to be the trajectory of a point corpuscle of mass *m* (a particle) rendering it in a time *t*, while the measuring unit in the kth scale is associated with the indetermination in the position of the corpuscle, then it is possible to demonstrate that when the indetermination of the corpuscle position diminishes, the indetermination in its linear momentum grows unboundedly. Based on the concept of similarity dimension of a corpuscle trajectory, from the before stated line of reasoning an alternative deduction of Heisenberg's uncertainty relation $\Delta x \Delta p_x \sim h$ is developed and discussed.

**Introduction**

The Heisenberg uncertainty relations (1925) [1] are considered one of the most fundamental results of quantum mechanics. We recall that their origin was associated with the duality point of view put forward by Louis de Broglie in 1924 [2]. It is interesting to note that one year before, W. Duane [3] interpreted the diffraction of X-rays in corpuscular terms (photons), invoking only a quantum rule associated with the momentum exchange of the photons with the periodic arrange of atoms in a crystal, following the quantum rules of Plank for the black body radiation [4] and of Bohr [5] for the motion of an electron around the nucleus of an atom.

Héctor Ceceña-Álvarez. Raúl W. Gómez-Gonzalez.
Facultad de Ciencias, Universidad Nacional Autónoma de México. Circuito Exterior, C. U., México D.F., 04510. México.
Antonio Peimbert-Torres. Instituto de Astronomía. Universidad Nacional Autónoma de México. Circuito Exterior, C. U., México D.F., 04510. México.

Ever since, different points of view of how Quantum Mechanics should be interpreted have been published –extensively treated in books [6-10 and references therein] and, until present, there is no answer that resolves the controversies in a definite way. In 1948 R. Feynman developed the path integral formalism [11, 12] in which he showed that one could interpret the path of a quantum particle as a continuous but not differentiable curve. A different approach was developed in D. Bohm's quantum theory [13, 14], in which quantum particles have precise trajectories. In 1981, Abbot y Wise showed that if the behaviour of a particle is subjected to quantum rules, the fractal dimension of its trajectory is two [15]. That such type of trajectory is plausible must be associated with the unpredictable and uncontrollable disturbance suffered by the physical system during a measurement. If this result can be made congruent with Heisenberg relations, then such relations imply trajectories with fractal dimension of two.

In this work we show an alternative way to arrive to uncertainty relations similar to those of Heisenberg in a way that does not invoke the duality of matter. Our starting point is to record the "trajectory" of a particle moving a certain distance in certain time. To do so, a series of cameras with different resolutions are used to record the movement. We make the fundamental assumption that any recording device is also a "quantum object" (that is, the recording devices are also subject to quantum rules) so that the process of recording (measurement) disturbs the original movement of the particle. The magnitude of this perturbation must be a function of the capability of the camera used to distinguish details of the "trajectory". The minimum perturbation possible is such that the exchange of action is of the order of $h$. We explore what happens if the resulting "trajectory" is a convergent self-similar curve. To simplify the problem we will illustrate the procedure with a Koch curve.

It is important to point out that the "trajectory" of a particle should not be taken in the classical sense; that is, the simultaneous and precise specification of its position and velocity (or momentum) for every instant of time. Rather, a geometric description of the "trajectory" will be established in terms of the scale (resolution) used and the changes of the areolar velocity (described below). In what follows, we will omit the inverted comas in the word trajectory.

**Methods**

We want to calculate the measure (length) of the trajectory as well as the velocity of a corpuscle of mass m traveling in a plane from region A to region B in a time t. With this objective in mind, a thought experiment is proposed in which a corpuscle is filmed k times, prepared under the same initial conditions, in k different scales of measurement.

Each and every camera (symbolically represented as "$C_k$") has a sub-index corresponding to the measuring scale each one of them records, its maximum resolution being the kth power of 3. In this way, the least resolution camera $C_0$ will not be able to resolve details (lengths) smaller than $\Delta x_0 = L_0$: camera $C_1$, being 3 times as powerful as camera $C_0$, will have a resolution:

$$\Delta x_1 = \frac{\Delta x_0}{3} = \frac{L_0}{3}. \tag{1}$$

The minimum length detectable by the kth camera $C_k$, its resolution being $3^k$ times that of $C_0$, is:

$$\Delta x_k = \frac{\Delta x_0}{3^k} = \frac{L_0}{3^k}. \tag{2}$$

These lengths will be, in turn, the units by which the trajectory will be measured in each and every scale, the kth measure taken as the product of the minimum detectable length $\Delta x_k$, times the number of cells $N_k$ visited by the corpuscle in that same scale.

**Uniform Rectilinear Motion**

A corpuscle goes from region A to region B following a rectilinear trajectory recorded by cameras $C_0$, $C_1$, $C_2$ … $C_k$. as described in the preceding section (figure 1). The trajectory's length in the kth scale is:

$$L_k = N_k \cdot \Delta x_k = 3^k \cdot \left(\frac{\Delta x_0}{3^k}\right) = L_0 \quad \forall k \in N \tag{3}$$

with a kth scale velocity being:

$$V_k = \frac{L_k}{\Delta t} = \frac{L_0}{\Delta t} = V_0 \quad \forall k \in N \tag{4}$$

Therefore, a particle undergoing uniform rectilinear motion describes a trajectory of length $L_0$ in each and every scale of measurement; that is to say, the trajectory's measure (its length) and its velocity are invariant under changes of scale.

**The Koch trajectory**

The same procedure is carried out with a Koch trajectory, as registered by cameras $C_0, C_1, C_2 \ldots C_k$ (figure 2). In this case $N_k = 4^k$ and the trajectory's length turns out to be:

$$L_k = N_k \cdot \Delta x_k = 4^k \cdot \left(\frac{\Delta x_0}{3^k}\right) = \left(\frac{4}{3}\right)^k L_0 \quad \forall k \in N$$

$$\Rightarrow L = \lim_{k \to \infty} \left(\frac{4}{3}\right)^k L_0 = \infty \tag{5}$$

and the kth velocity:

$$v_k = \frac{L_k}{\Delta t} = \left(\frac{4}{3}\right)^k \frac{L_0}{\Delta t}$$

$$\Rightarrow V = \lim_{k \to \infty} v_k = \infty. \tag{6}$$

Differently from the first case, in the Koch curve the trajectory's length and corpuscle's velocity both diverge when the measuring scale decreases; that is to say, they are not invariant against changes of scale.

**Area as a measure of the trajectory**

If instead of trying to find a length as a measure of the trajectory, an attempt is made to find it by means of its area, the outcome is no better than the one already found:

$$A_k = N_k \cdot (\Delta x_k)^2 = 4^k \cdot \left(\frac{\Delta x_0}{3^k}\right)^2 = \left(\frac{4}{9}\right)^k L_o^2 \quad \forall k \in N$$

$$\Rightarrow A = \lim_{k \to \infty} \left(\frac{4}{9}\right)^k L_o^2 = 0$$

(7)

When the measuring scale decreases, the area of the trajectory tends to zero

**Analysis extension**

The analysis just carried out can be extended to self-similar forms in bi-dimensional space through the relations:

$$\Delta x_k = \frac{\Delta x_0}{\rho^k} \quad ; \quad L_k = N_k \cdot \Delta x_k \quad ; \quad A_k = N_k \cdot (\Delta x)^2.$$

(8)

Due to self-similarity, it holds true that:

$$N_k = N^k \quad ; \quad D_s = \log_\rho N \Leftrightarrow N = \rho^{D_s}$$

(9)

and the k-th area can be expressed as:

$$A_k = N_k \cdot (\Delta x_k)^2 = \Delta x_k \cdot L_k = L_0^2 \cdot \rho^{k(D_s-2)}. \tag{10}$$

Subtracting from $A_k$ the quantity

$$A_{k0} = \Delta x_k \cdot L_0 = L_0^2 \cdot \rho^{-k} \tag{11}$$

we obtain:

$$\Delta A_{k0} = \Delta x_k \cdot L_k - \Delta x_k \cdot L_0 = \Delta x_k \cdot \Delta L_k = L_0^2 \cdot \left\{ \rho^{k(D_s-2)} - \rho^{-k} \right\}$$

$$\Delta A_{k0} = L_0^2 \cdot \gamma(k, \rho, D_s) \quad ; \quad \gamma(k, \rho, D_s) = \left\{ \rho^{k(D_s-2)} - \rho^{-k} \right\} \tag{12}$$

where $\Delta A_{k0}$ is defined as the kth change in the trajectory surface with respect to scale. Even though $\gamma(k, \rho, D_s)$ is formally a three variable function, when the analysis for a specific trajectory is undertaken, $\rho$ and $D_s$ play the role of parameters while $\gamma = \gamma(k)$ is actually a function only of k, with upper and lower bounds given in terms of the similarity dimension of the trajectory. At this point, it is possible to demonstrate that:

$$\text{If } D_s > 2 \quad \Rightarrow \quad \infty > \Delta x_k \cdot \Delta L_k > \frac{L_0^2}{2}. \tag{13}$$

$$\text{If } D_s = 2 \quad \Rightarrow \quad L_0^2 > \Delta x_k \cdot \Delta L_k \geq \frac{L_0^2}{2}. \tag{14}$$

$$\text{If } 2 > D_s > 1 \quad \Rightarrow \quad L_0^2 > \Delta x_k \cdot \Delta L_k > 0. \tag{15}$$

$$\text{If } D_s = 1 \quad \Rightarrow \quad \Delta x_k \cdot \Delta L_k = 0. \tag{16}$$

If the particle's mass m and its traveling time $\Delta t$ are considered to be invariant under changes of scale, it is possible to further derive:

$$\frac{1}{\Delta t} \cdot A_k = \frac{\Delta x_k \cdot \Delta L_k}{\Delta t} = \Delta x_k \cdot \Delta v_k = \Delta V_k$$

$$\Rightarrow m \cdot \Delta V_k = \Delta x_k \cdot \Delta p_k = \Delta P_k \tag{17}$$

defined to be the kth change in areolar velocity with respect to scale and the kth change in the areolar momentum with respect to scale, respectively. But since $\Delta A_k = L_k^2 \cdot \gamma(\rho, D_s)$ (Eq 12), then

$$\frac{1}{\Delta t} \cdot A_k = \Delta x_k \cdot \Delta v_k = m \frac{\Delta x_k \cdot \Delta L_k}{\Delta t} = m \frac{L_0^2 \gamma(\rho, D_s)}{\Delta t} = 2E_0 \gamma(\rho, D_s) \Delta t = 2\eta_0 \tag{18}$$

Being so, the four relations given above (13 to 16) take the form:

$$\text{If } D_s > 2 \quad \Rightarrow \quad \infty > \Delta x_k \cdot \Delta p_k > \eta_0. \tag{19}$$

$$\text{If } D_s = 2 \quad \Rightarrow \quad 2\eta_0 > \Delta x_k \cdot \Delta p_k \geq \eta_0. \tag{20}$$

$$\text{If } 2 > D_s > 1 \quad \Rightarrow \quad 2\eta_0 > \Delta x_k \cdot \Delta p_k > 0. \tag{21}$$

$$\text{If } D_s = 1 \quad \Rightarrow \quad \Delta x_k \cdot \Delta p_k = 0 \tag{22}$$

where $\eta_0 = E_0 \Delta t$ has dimensions of action.

**Results and discussion**

Because $\eta_0$ has the dimensions of action, one is tempted to identify it with Plack's constant $h$, so the right hand of this equation is formally equivalent to Heisenberg's uncertainty relation $\Delta x \Delta p_x \geq h$ for the case that $D_s = 2$. However, the left side of Eq. (20) establishes an upper limit to the uncertainty relations. This result does not seem unlikely if one thinks that the exchange of action during the measurement is one quantum of action at a time.

When the measurement procedure does not involve a great perturbation in the particle state, that is, if during the time interval $\Delta t$ elapsed in the measurement the fractional change in the particle's energy is small, the fractal features of the trajectory become less noticeable and the dimension of similarity tends to one, in accordance with Bohr's correspondence principle; that is, $D_s = 1$ then $\Delta x \Delta p_x = 0$. Actually, when $2 > D_s > 1$, our results still give a weaker form of uncertainty relations that could be associated with the transition from quantum to classical mechanics.

**Conclusions**

The self-similar description herein described is a corpuscular construction of an indetermination relation between position and momentum that:

a) Provides an alternative geometrical (Eqs. 13 to 16) and physical (Eqs. 18 to 21) meaning to the uncertainty relation between the position and the linear momentum of a corpuscle, not previously reported.

b) Does not make use of any wave-like hypothesis whatsoever to reach a conclusion originally conceived from assuming the dual character of quantum objects.

c) Provides a general formulation for uncertainty relations between linear momentum and position of a corpuscle as a function of the similarity dimension of the trajectory from which we obtain:

 A quantization rule for the linear momentum when $D_s = 2$, from which uncertainty relations quite similar to those proposed by W. Heisenberg in 1927 are derived.

d) Endow a relation compatible with Bohr's correspondence principle ($\eta \to 0$) and a conservation law for linear momentum when $D_s = 1$.

e) Suggests that the transition from quantum to classical mechanics is related to the dimension of similarity of the trajectory.

Figure captions.

Figure 1. Rectilinear trajectory of a corpuscle as registered by cameras $C_0$, $C_1$, $C_2$.

Figure 2. Koch trajectory of a corpuscle as registered by cameras $C_0$, $C_1$, $C_2$

Figure 1

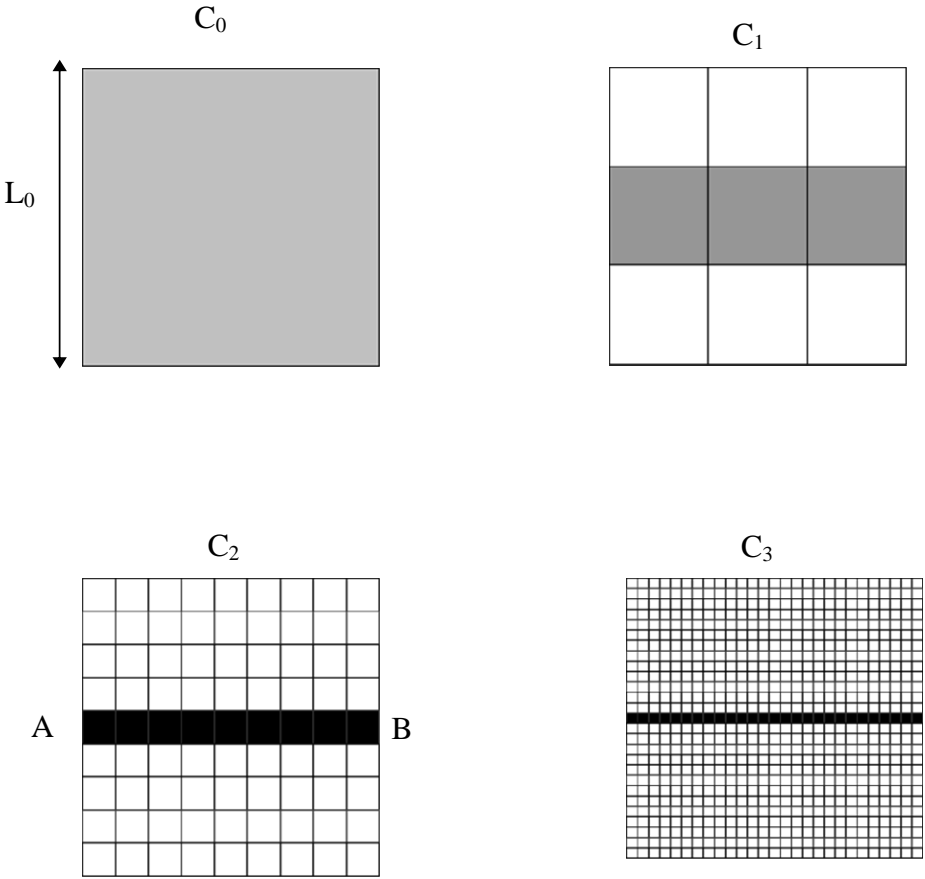

Figure 2.

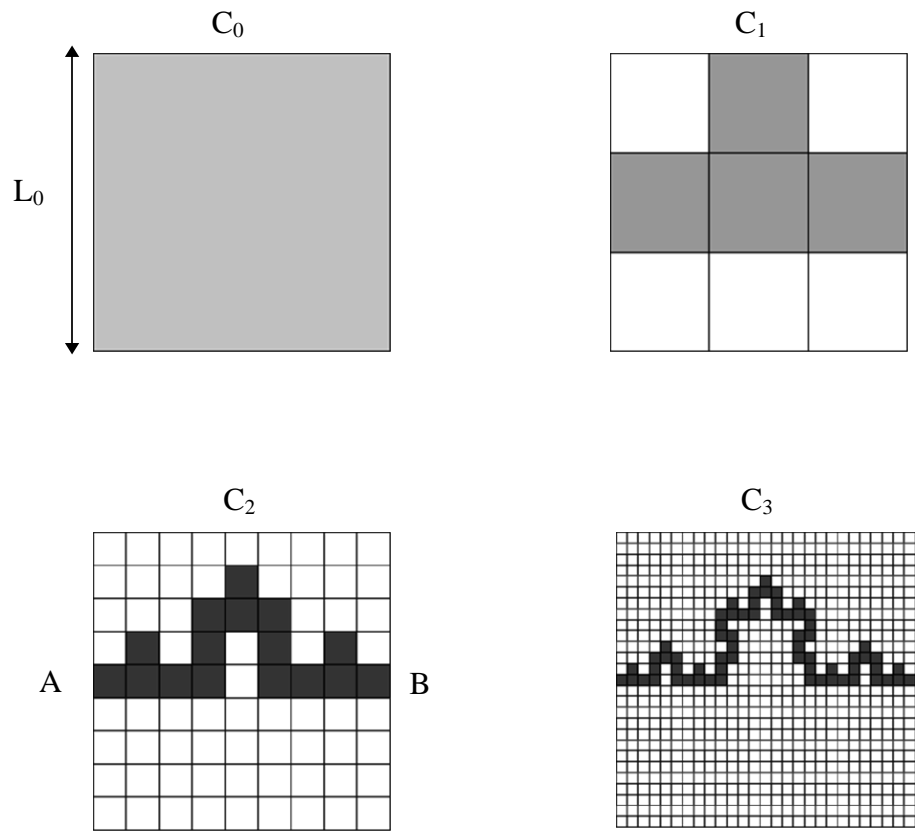